\documentclass[twocolumn,aps,prl,superscriptaddress,letterpaper]{revtex4}

\usepackage{graphicx}
\usepackage{dcolumn}
\usepackage{bm}
\usepackage{amsmath, amssymb}%
\usepackage{epstopdf}
\usepackage{color}
\usepackage{soul}
\usepackage{hhline}
\usepackage{multirow}
\usepackage[usestackEOL]{stackengine}

\setstackgap{L}{\normalbaselineskip}

\usepackage{float}

\newcommand{\prel}{\ifmmode p_{rel} \else $p_{rel}$ \fi}
\newcommand{\pcm}{\ifmmode p_{cm} \else $p_{cm}$ \fi}

\usepackage[]{changes} 

\begin{document}

\title{Neutron valence structure from nuclear deep inelastic scattering}

\newcommand*{\MIT }{Massachusetts Institute of Technology, Cambridge, Massachusetts 02139, USA}
\newcommand*{\ODU}{Old Dominion University, Norfolk, Virginia 23529, USA}
\newcommand*{\JLAB}{Thomas Jefferson National Accelerator Facility, Newport News, Virginia 23606, USA}
\newcommand*{\TAU }{School of Physics and Astronomy, Tel Aviv University, Tel Aviv 69978, Israel}
\newcommand*{\Penn}{Pennsylvania State University, University Park, PA, 16802, USA}
\newcommand*{\GW}{George Washington University, Washington, D.C., 20052, USA}

\author{E.P. Segarra}
\affiliation{\MIT}
\author{A. Schmidt}
\affiliation{\MIT}
\affiliation{\GW}
\author{T. Kutz}
\affiliation{\MIT}
\affiliation{\GW}
\author{D.W. Higinbotham}
\affiliation{\JLAB}
\author{E.~Piasetzky}
\affiliation{\TAU}
\author{M. Strikman}
\affiliation{\Penn}
\author{L.B. Weinstein}
\affiliation{\ODU}
\author{O. Hen}
\email[Contact Author \ ]{hen@mit.edu}
\affiliation{\MIT}

\date{\today{}}

\begin{abstract}

  Mechanisms of spin-flavor SU(6) symmetry breaking in Quantum Chromodynamics
  (QCD) are studied via an extraction of the free neutron structure
  function from a global analysis of deep inelastic scattering
  (DIS) data on the proton and on nuclei from $A = 2$ (deuterium) to
  208 (lead). Modification of the structure function of nucleons bound
  in atomic nuclei (known as the EMC effect) are consistently accounted for
  within the framework of a universal modification of nucleons in
  short-range correlated (SRC) pairs.
  Our extracted neutron-to-proton structure function ratio
  $F_2^n/F_2^p$ becomes constant for $x_B \ge 0.6$, equalling $0.47
  \pm 0.04$ as $x_B \rightarrow 1$, in agreement with theoretical
  predictions of perturbative QCD and the Dyson Schwinger equation,
  and in disagreement with predictions of the Scalar Diquark dominance
  model.  We also predict $F_2^{^3\mathrm{He}}/F_2^{^3\mathrm{H}}$,
  recently measured, yet unpublished, by the MARATHON collaboration, the nuclear
  correction function that is needed to extract $F_2^n/F_2^p$ from
  $F_2^{^3\mathrm{He}}/F_2^{^3\mathrm{H}}$, and the theoretical
  uncertainty associated with this extraction.

\end{abstract}

\maketitle


\section{Introduction}
Almost all the visible mass in the universe comes from the mass of
protons and neutrons, and is dynamically generated by the strong
interactions of quarks and gluons~\cite{Bashir:2012}. These
interactions are described by the theory of strong interactions,
Quantum Chromodynamics (QCD). While the structure of low-energy QCD
largely follows spin-flavor SU(6) symmetry, this symmetry is broken, as evident by the
mass difference between the proton and its first excited state, the
Delta resonance. The exact symmetry-breaking mechanism is still an
open question.  This affects our understanding of emergent QCD phenomena
such as baryon structure, masses, and magnetic
moments~\cite{Roberts:2013mja}. Answering this question is thus one of the
main motivations for the ongoing international effort to measure the
quark-gluon structure of hadrons.

Different symmetry-breaking mechanisms can be discriminated among
experimentally by measuring nucleon structure functions, which are
sensitive to the distributions of quarks inside nucleons.
Specifically, realistic models of QCD make very different predictions
for the relative probability for a single quark to carry all of the
momentum of a neutron compared to that of a proton, i.e., the proton
to neutron structure function ratio, $F_2^n(x_B,Q^2)/F_2^p(x_B,Q^2)$,
as $x_B\rightarrow 1$ (where $x_B = Q^2/2m\nu$ is the fractional quark
momenta in the collinear reference frame where the nucleon is fast, 
$Q^2$ is the four-momentum transfer squared, $m$ is the
nucleon mass, and $\nu$ is the energy transfer).

While the proton structure function has been extensively measured, the
lack of a free neutron target prevents equivalent measurements of the
neutron structure function, thereby preventing a direct test of QCD
symmetry breaking mechanisms. 

Here we use measurements of all available structure functions of nuclei  
(ranging from deuterium to lead) to extract the free neutron structure function,
while consistently accounting for the nuclear-medium induced
modification of the quark distributions in atomic nuclei. Using data
on such a wide span of nuclei provides a large lever arm that allows
us to precisely constrain $F_2^n(x_B,Q^2)/F_2^p(x_B,Q^2)$, obtaining
new insight into the fundamental structure of QCD.

We find that as $x_B$ approaches unity,
$F_2^n(x_B,Q^2)/F_2^p(x_B,Q^2)$ saturates at a value of $0.47 \pm
0.04$, giving credence to modern predictions of QCD such as those
based on the Dyson Schwinger Equation ($0.41-0.49$)~\cite{Roberts:2013mja} and
Perturbative QCD ($3/7$)~\cite{Farrar:1975yb}. This contrasts with
previous extractions that did not include DIS measurements of nuclei
heavier than deuterium~\cite{Dulat:2015mca,Accardi:2016qay,Arrington:2011qt} and 
claimed to support the scalar di-quark ($1/4$)~\cite{Close:1973xw,
  Carlitz:1975bg} view of the nucleon.

The large differences between previous extractions of
$F_2^n(x_B,Q^2)/F_2^p(x_B,Q^2)$ and those of this work emphasize the
need for direct experimental verification. The MARATHON
Experiment~\cite{MARATHON} recently measured
$F_2^{^3\mathrm{He}}(x_B,Q^2)/F_2^{^3\mathrm{H}}(x_B,Q^2)$ with the
goal of providing an independent determination of
$F_2^n(x_B,Q^2)/F_2^p(x_B,Q^2)$ with minimal sensitivity to 
nuclear medium effects. This extraction is based on the assumption
that such effects should be very similar for ${^3\mathrm{He}}$ and
${^3\mathrm{H}}$, thereby cancelling in their ratio. Using the results
of our global analysis, we present predictions for the
$F_2^{^3\mathrm{He}}(x_B,Q^2)/F_2^{^3\mathrm{H}}(x_B,Q^2)$ ratio and
the nuclear correction function required to extract
$F_2^n(x_B,Q^2)/F_2^p(x_B,Q^2)$ from it. By comparing our correction
function with those of earlier works we quantify the model uncertainty
associated with this extraction, which can be as high as $\sim25\%$
for current realistic models.

\section{Universal Nucleon Modification and the EMC Effect}
Given the lack of a free neutron target,
the modification of the quark-gluon structure of nucleons bound in
atomic nuclei, known as the EMC effect, is the main issue preventing a
direct extraction of the free neutron structure function from lepton
Deep Inelastic Scattering (DIS) measurements of atomic nuclei, see
Ref.~\cite{Hen:2016kwk} for a recent review.

We account for the EMC effect in nuclear DIS data by exploiting recent
insight to its origin, gained from observations of a
correlation between the magnitude of the EMC effect in different
nuclei and the relative amount of short-range correlated (SRC) nucleon
pairs in those nuclei~\cite{weinstein11, Hen12, Hen:2013oha,
  Hen:2016kwk, Schmookler:2019nvf, Frankfurt88}.

SRC pairs are predominantly proton-neutron ($pn$) pairs~\cite{piasetzky06,
  subedi08, korover14, hen14, duer18, Duer:2018sxh}.  They have large
relative and individual momenta, smaller center-of-mass momenta, and
account for 60-70\% of the kinetic energy carried by nucleons in the
nucleus~\cite{tang03,shneor07,korover14,Cohen:2018gzh}.  Therefore,
nucleons in such pairs have significant spatial overlap and are far
off their mass-shell ($E^2-p^2-m^2<0$).

These extreme conditions, and the observed correlation between SRC pair
abundances and the magnitude of the EMC effect, imply that the EMC
effect could be driven primarily by the modification of the structure
functions of nucleons in SRC pairs~\cite{weinstein11, Hen12,
  Hen:2016kwk}.

Utilizing scale separation between SRC and uncorrelated (mean-field)
nucleons, Ref.~\cite{Schmookler:2019nvf} modeled the nuclear structure
function as having contributions from unmodified uncorrelated nucleons
and from modified correlated nucleons in $np$-SRC pairs:
\begin{eqnarray}
\begin{split}
F_2^A = Z F_2^p + N F_2^n + n_{SRC}^A (\Delta F_2^{p} + \Delta F_2^{n}),
\label{Eq:model_f2a}
\end{split}
\end{eqnarray}
where $N$ and $Z$ are the number of neutrons and protons in the
nucleus ($N+Z=A$), $n_{SRC}^A$ is the average number of nucleons in
$np$-SRC pairs, $\Delta F_2^p$ and $\Delta F_2^n$ are the average
differences between the structure functions of free nucleons and
nucleons in SRC pairs, and we omitted the explicit $x_B$ and $Q^2$
dependence of the $F_2$ structure functions for brevity.  This model
assumes that both the EMC effect at $0.3\le x_B\le 0.7$ and 
nucleon-motion effects (which are important at $x_B>0.7$) are dominated by
short-range correlations~\cite{Frankfurt88,Sargsian02,Melnitchouk:1997pemc}.  Therefore
both are approximately proportional to SRC pair abundances and captured by
Eq.~\ref{Eq:model_f2a}.  This model neglects the contribution of $pp$- and $nn$-SRC
pairs that, due to the predominance of the Tensor interaction at short-distance,
are only $\approx10\%$ of all $NN$-SRC pairs in both light and heavy nuclei
~\cite{piasetzky06, subedi08, korover14, hen14, duer18, Duer:2018sxh},
and have little impact on our results.
See supplementary materials for details.

To reduce sensitivity to isospin, target-mass, and higher twist
effects~\cite{Virchaux:1991jc}, DIS data are traditionally given in
the form of $F_2^A/F_2^d$ ratios. We use Eq.~\ref{Eq:model_f2a} to
express this ratio as:
\begin{eqnarray}
\begin{split}
\frac{F_2^A}{F_2^d} =& & \frac{\Delta F_2^{p} + \Delta F_2^{n}}{F_2^d/n_{SRC}^d} &\times (\frac{n_{SRC}^A}{n_{SRC}^d}-N) + 
(Z-N)\frac{F_2^p}{F_2^d}+N \\
=& &f_\text{univ}(x_B) &\times (\frac{n_{SRC}^A}{n_{SRC}^d}-N) + (Z-N)\frac{F_2^p}{F_2^d}+N,
\label{Eq:model_emc}
\end{split}
\end{eqnarray}
where we defined a nucleus independent universal modification function (UMF)
\begin{eqnarray}
\begin{split}
f_\text{univ} = n_{SRC}^d\frac{\Delta F_2^{p} + \Delta F_2^{n}}{F_2^d}.
\label{Eq:f_univ}
\end{split}
\end{eqnarray}

\begin{figure} [t]
\includegraphics[width=0.9\columnwidth, height=5.5cm]{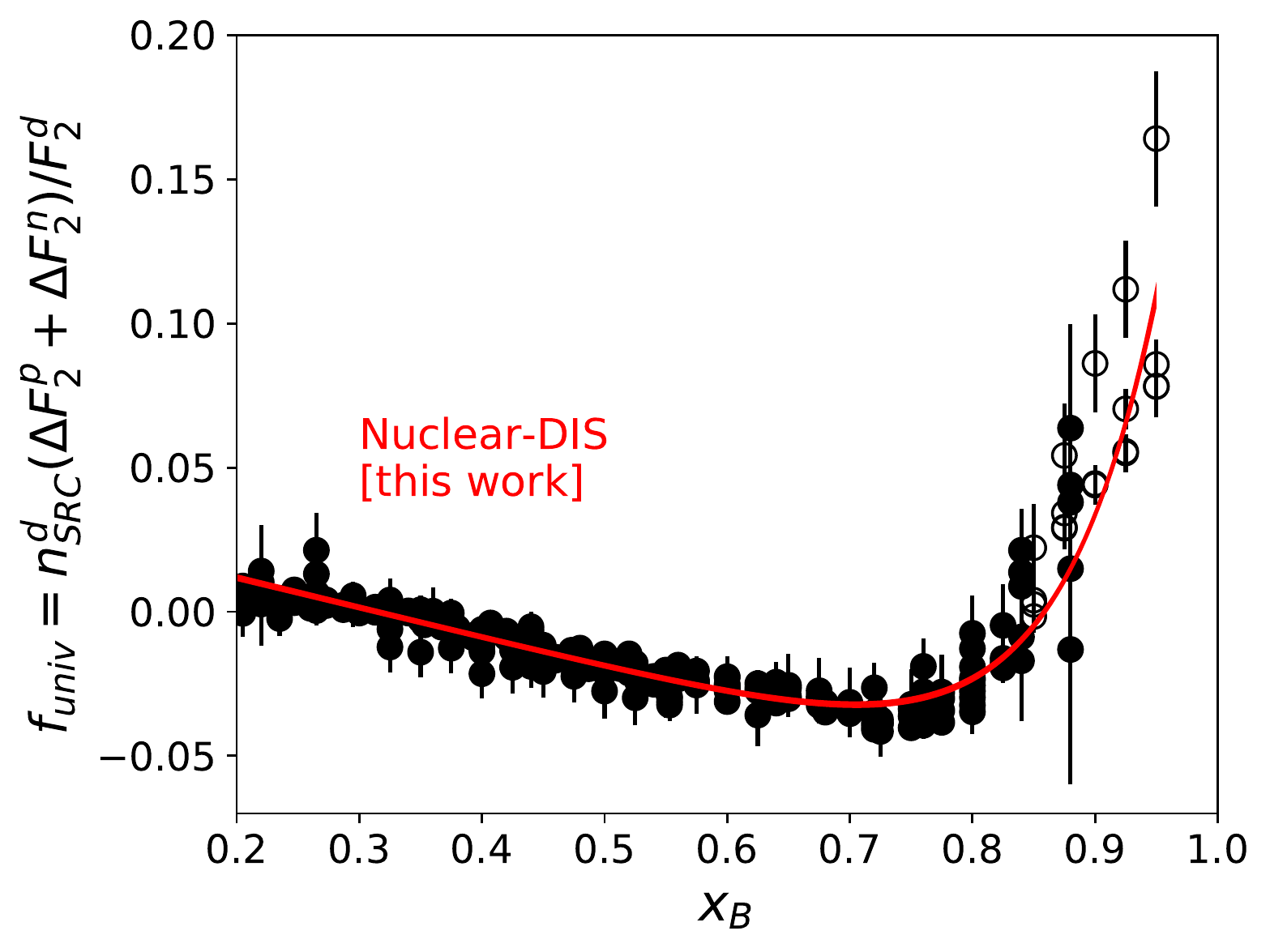}
\caption{The extracted global universal modification function (UMF) from the
  Nuclear-EMC effect analysis performed here (red band). The narrow width of the band
  shows the 68\% confidence interval. Data points show the data-driven
  extractions of Ref.~\cite{Schmookler:2019nvf}, based on individual
  measurements of $F_2^A/F_2^d$ in a variety of nuclei. Open and
  closed data points show measurements at $W<\sqrt{2}$ GeV and $W\ge \sqrt{2}$ GeV respectively.}
\label{Fig:uniFunct}
\end{figure}

Consistent UMFs were previously extracted for nuclei from $^3$He to
$^{208}$Pb, pointing to the existence of a global UMF for SRC pairs in
any nucleus (see
Fig.~\ref{Fig:uniFunct})~\cite{Schmookler:2019nvf}. Here we extract
the global UMF using Bayesian inference by means of a Hamiltonian
Markov Chain Monte Carlo (HMCMC)~\cite{JSSv076i01,pystan}, referred to
herein as Nuclear-DIS analysis.

We parametrized the UMF for all nuclei as 
\begin{eqnarray}
\begin{split}
f_\text{univ} = \alpha + \beta x_B + \gamma e^{\delta(1-x_B) }
\label{eq:umfform}
\end{split}
\end{eqnarray}
and estimated its parameters ($\alpha$, $\beta$, $\gamma$, and
$\delta$) using HMCMC-based inference from $F_2^A/F_2^d$
data~\cite{Schmookler:2019nvf,Gomez94,Seely:2009gt} for $0.08\le
x_B\le 0.95$ in $^3$He, $^4$He, $^9$Be, $^{12}$C, $^{27}$Al,
$^{56}$Fe, $^{197}$Au, and $^{208}$Pb, via Eq.~\ref{Eq:model_emc}.
Here, and throughout this work, we consistently removed all isoscalar
corrections previously applied to asymmetric nuclei data.  We assumed
$\frac{n_{SRC}^A/A}{n_{SRC}^d/2} = a_2(A/d)$, the average per-nucleon
cross-section ratio for quasi-elastic electron scattering in nucleus
$A$ relative to deuterium at
$1.5<x_B<2$~\cite{frankfurt93,egiyan02,egiyan06,fomin12,Hen12,Schmookler:2019nvf}.
$F_2^p/F_2^d$ is taken from Table~2 of
Ref.~\cite{Arrington:2008zh}. As consistent parameterizations of
$F_2^p/F_2^d$ as a function of $x_B$ are needed for the UMF
extraction, we parametrized it as $F_2^p / F_2^d= \alpha_d + \beta_d
x_B + \gamma_d e^{\delta_d(1-x_B) }$. We determine all parameters, including those of the UMF
and $F_2^p / F_2^d$ simultaneously from data as part of the Nuclear-DIS analysis. See online
supplementary materials for details on the inference procedure,
posterior distributions of the parameters, and discussion of the
kinematical coverage of the fitted data.

The Nuclear-DIS analysis reproduced all the $F_2^A/F_2^d$ data over
the entire measured $x_B$ range, see online supplementary
materials Fig. 1. The resulting global UMF (red band in Fig.~\ref{Fig:uniFunct}) extends
up to $x_B\sim0.95$ and agrees well with the individual nuclear UMFs extracted in Ref~\cite{Schmookler:2019nvf}.


\section{$F_2^n/F_2^p$ Extraction}

\begin{figure} [t]
\includegraphics[width=\columnwidth]{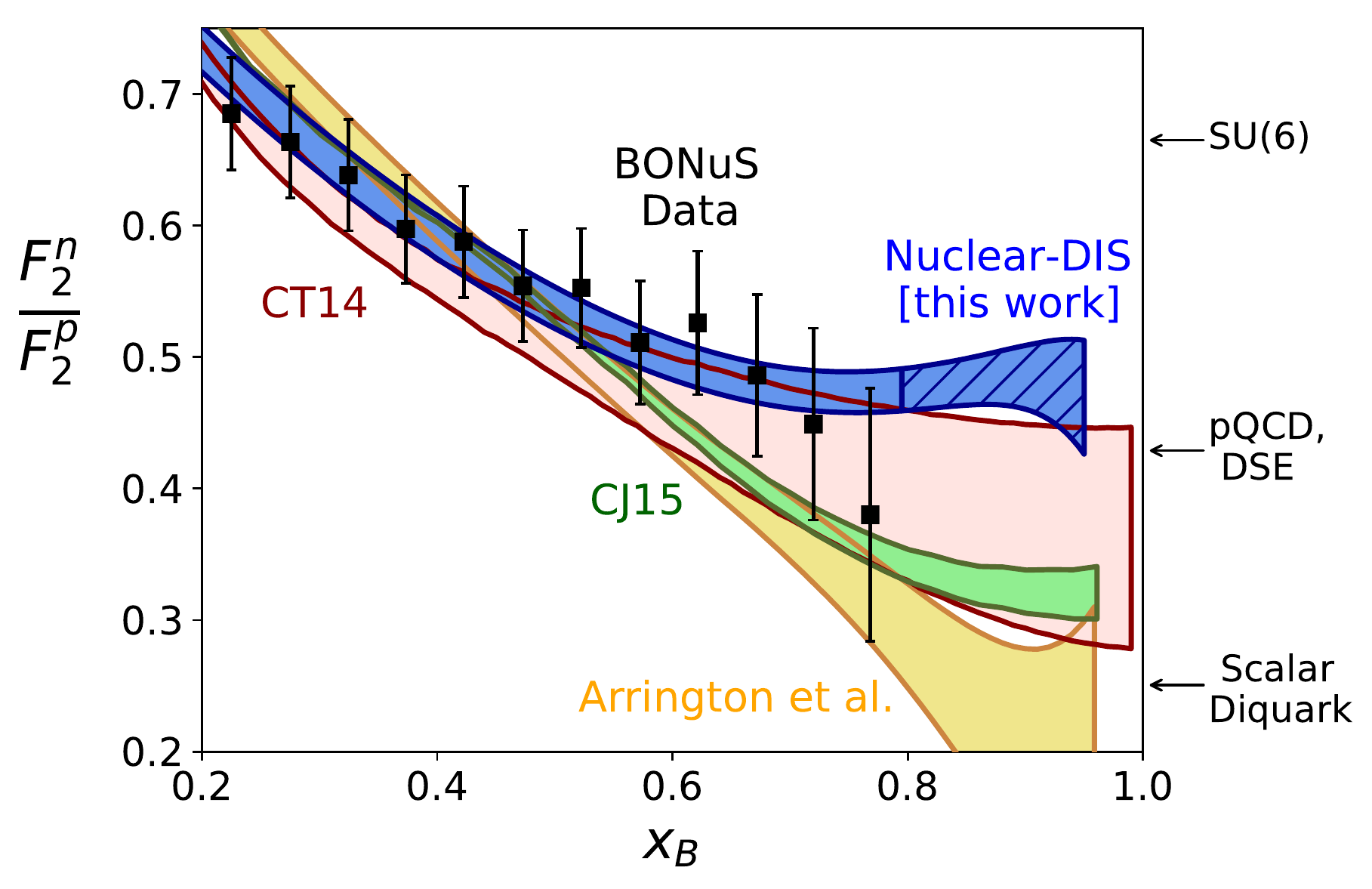}
\caption{Neutron-to-proton structure function ratio
  $F_2^n/F_2^p$. Data points show the $d(e,e'p_S)$ tagged-DIS
  measurement~\cite{Baillie:2011za}. Our predictions (blue band labeled
  `Nuclear-DIS', including a 68\% confidence interval) are compared with those of
  CT14~\cite{Dulat:2015mca} (red band), 
  CJ15~\cite{Accardi:2016qay} (green band), and Arrington et
  al.~\cite{Arrington:2011qt} (yellow band), which 
  treat nuclear effects in deuterium DIS data differently (see text for
  details).  The labels show $F_2^n/F_2^p$ predictions at $x_B=1$, such
  as SU(6) symmetry~\cite{Close:1979bt}, perturbative QCD
  (pQCD)~\cite{Farrar:1975yb}, Dyson-Schwinger Equation
  (DSE)~\cite{Roberts:2013mja} and Scalar Diquark
  models~\cite{Close:1973xw, Carlitz:1975bg}. All predictions are
  obtained within the parton model framework~\cite{Virchaux:1991jc}
  and all extractions were consistently evolved to the same value of
  $Q^2$ based on the kinematics of the MARATHON
  experiment~\cite{MARATHON}, i.e. $Q^2 = (14 \mathrm{~GeV}^2) \times
  x_B$. }
\label{Fig:f2nf2p}
\end{figure}

Using Eq.~\ref{Eq:model_f2a} to model nuclear effects in $F_2^d$ we express $F_2^n/F_2^p$ as:
\begin{eqnarray}
\begin{split}
\frac{F_2^n}{F_2^p} = \frac{1 - f_\text{univ}}{F_2^p/F_2^d} - 1.
\label{Eq:model_f2nf2p}
\end{split}
\end{eqnarray}
We extract $F_2^n/F_2^p$ using $f_\text{univ}$ and $F_2^p/F_2^d$
determined by our Nuclear-DIS analysis discussed above (see
Fig.~\ref{Fig:f2nf2p}).  Our results are consistent with the
experimental extraction using tagged $d(e,e'p_S)$ DIS measurements on
the deuteron~\cite{Baillie:2011za}.  $F_2^n/F_2^p$ decreases steadily for $0.2 \le x_B
< 0.6$, and becomes approximately constant starting at $x_B \approx
0.6$. The $x_B \rightarrow 1$ limit of $F_2^n/F_2^p$ equals $0.47 \pm
0.04$.

Removing low-$W$ DIS data ($W < \sqrt{2}$ GeV) from our analysis limits
our extraction to $x_B \sim 0.8$ but does not change its conclusions
since $F_2^n/F_2^p$ still saturates starting at $x_B \approx
0.6$. The hatched region of the blue band in Fig.~\ref{Fig:f2nf2p} corresponds to our model extraction using the low-$W$ DIS data to reach up to $x_B\sim0.95$. Similarly, we verified that evolving $F_2^p/F_2^d$ from
$Q_0^2=12$~GeV$^2$ to $Q^2 = 5$~GeV$^2$ does not significantly change
our extraction up to $x_B \sim 0.8$. See online supplementary
materials for details.

Our Nuclear-DIS analysis gives significantly larger values of $F_2^n/F_2^p$ than
several previous extractions which do not use $A>2$ nuclear-DIS data,
including: (A) CTEQ global analysis (CT14)~\cite{Dulat:2015mca}, which
uses $W$ ($> 3.5$ GeV) and $Q^2$ ($> 2$ GeV$^2$) DIS data
for $A\leq2$ (with no corrections for any nuclear effects in the
deuteron) combined with various other reactions such as jet production 
and $W^\pm,Z$ production, (B) CTEQ-JLab global analysis
(CJ15)~\cite{Accardi:2016qay}, which uses $A\leq2$ DIS data with
looser cuts of $W>1.7$ GeV and $Q^2>1.3$ GeV$^2$, together with
recently published $W^\pm$-boson charge asymmetries from
D0~\cite{D0:2014kma} and additional corrections for
deuterium off-shell, higher-twist, and target-mass effects, and (C)
Arrington et al.~\cite{Arrington:2011qt}, which includes only $A\leq2$
DIS data with only corrections for Fermi motion and binding (see
Fig.~\ref{Fig:f2nf2p}).

\begin{figure} [t]
\includegraphics[width=0.9\columnwidth]{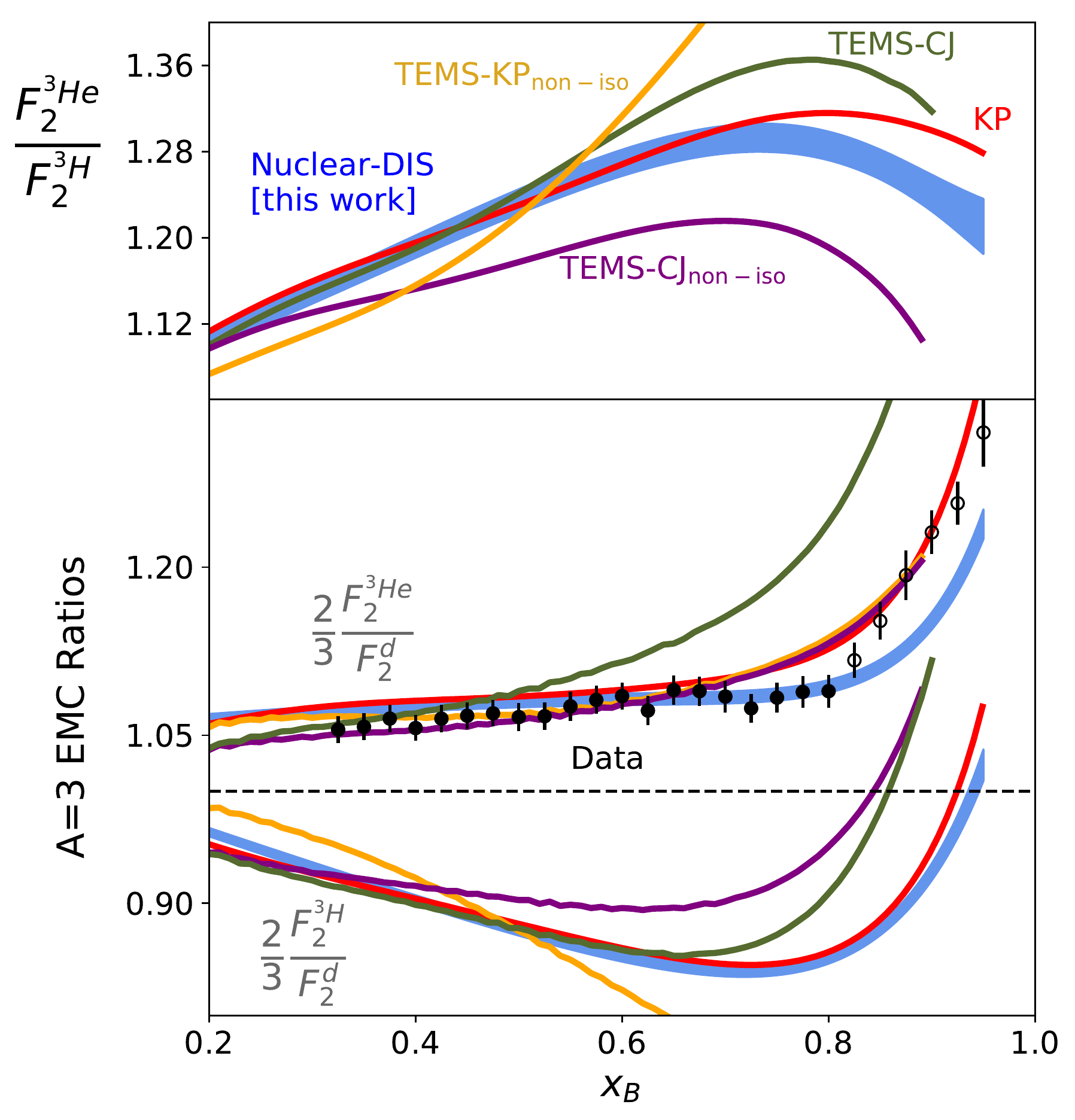}
\caption{Top: Nuclear-DIS analysis result for
  $F_2^{^3\mathrm{He}}/F_2^{^3\mathrm{H}}$ (blue band) compared to
  previous extractions by Tropiano et al.~\cite{Tropiano:2019tems}
  (TEMS [green, purple, orange, assuming different off-shell
  corrections]) and Kulagin and
  Petti~\cite{Kulagin:2019pc,Kulagin:2010gd} (KP [red]).  Bottom:
  Nuclear-DIS analysis results for $2F_2^{^3\mathrm{He}}/3F_2^d$,
  $2F_2^{^3\mathrm{H}}/3F_2^d$ shown in blue. The width of the bands
  show the 68\% confidence intervals of our analysis. Predictions for
  $2F_2^{^3\mathrm{H}}/3F_2^d$ are based on the assumption
  $n_{SRC}^{^3\mathrm{H}} = n_{SRC}^{^3\mathrm{He}}$. Symbols show the
  $2F_2^{^3\mathrm{He}}/3F_2^d$ measurement of
  Ref.~\cite{Seely:2009gt} re-scaled by $\sim$2\%. TEMS and KP lines
  do not include uncertainties. See text for details.}
\label{Fig:emcA3}
\end{figure}

\begin{figure*} [t]
\includegraphics[width=0.45\textwidth]{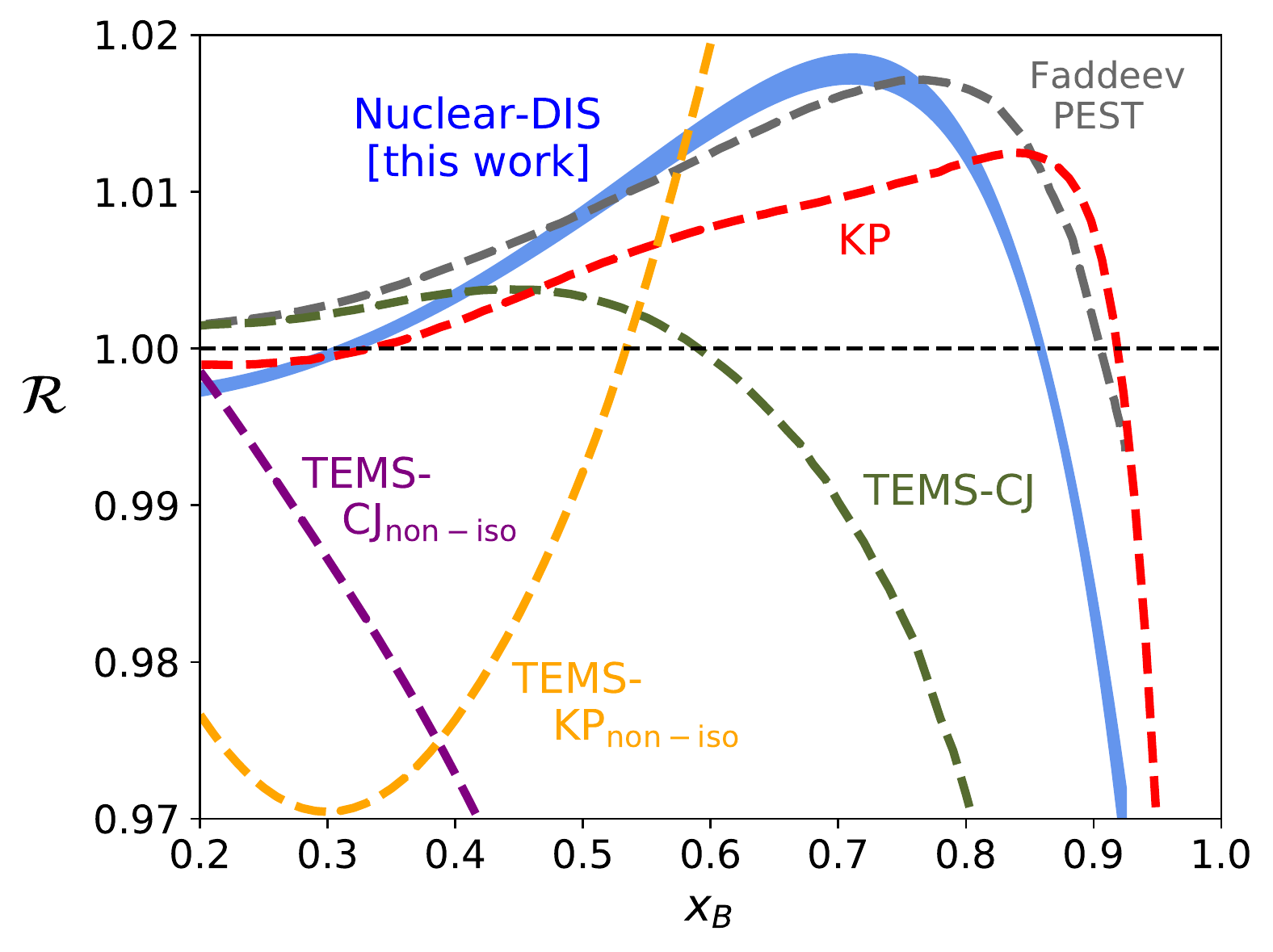}
\includegraphics[width=0.505\textwidth,height=6.01cm]{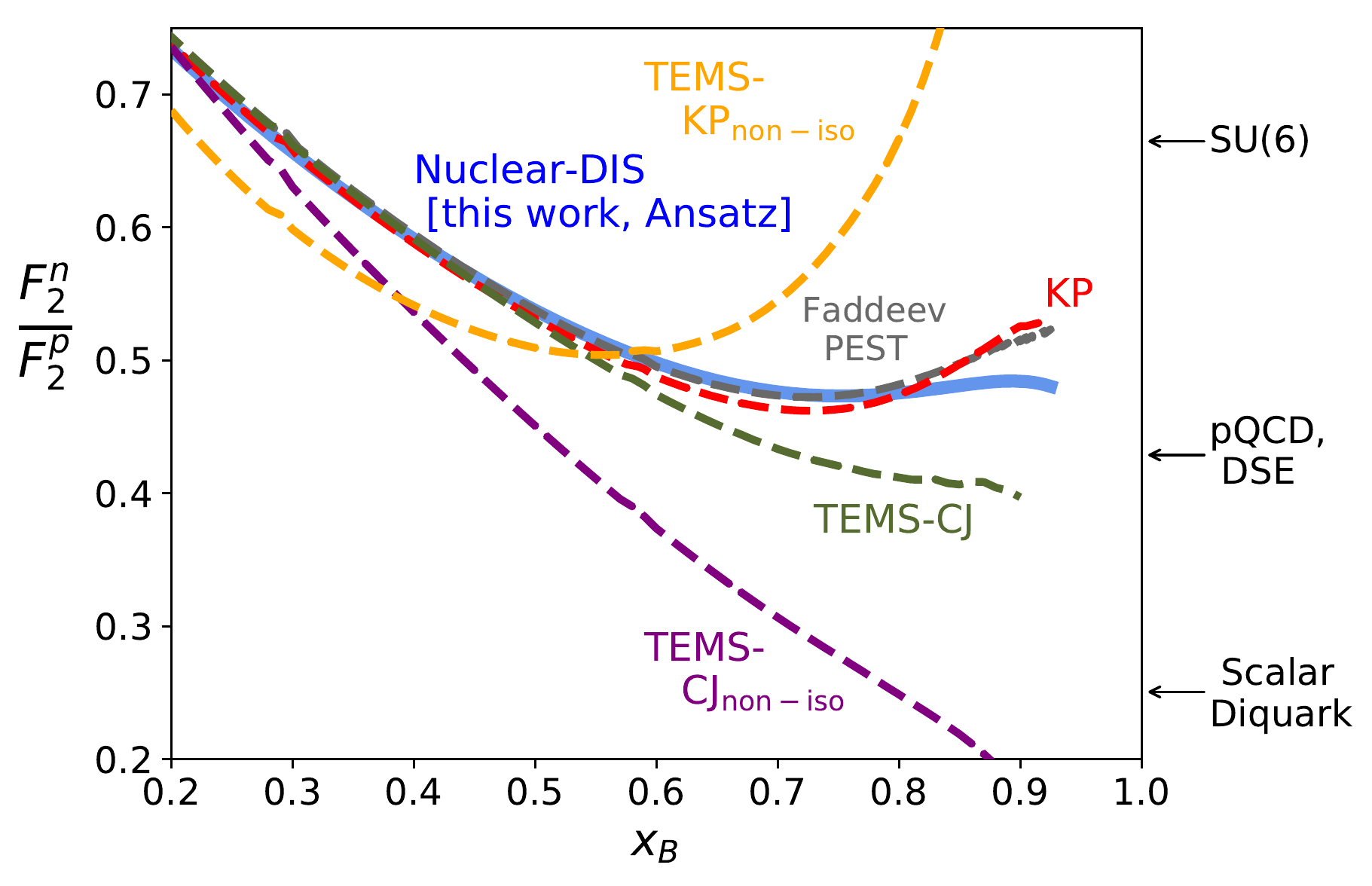}
\caption{Left: Different model predictions for $\mathcal{R}$ as
  defined in Eq.~\ref{Eq:R}. The present work (blue band, labeled
  `Nuclear-DIS') includes a 68\% confidence interval.  The other
  labeled lines show calculations (not including uncertainties) by Tropiano et
  al.~\cite{Tropiano:2019tems} (TEMS, for different off-shell
  corrections), Kulagin and Petti~\cite{Kulagin:2010gd,Kulagin:2019pc}
  (KP), and Afnan et al.~\cite{Afnan:2003vh} (Faddeev PEST). Right:
  $F_2^n/F_2^p$ extracted from predicted values of
  $F_2^{^3\mathrm{He}}/F_2^{^3\mathrm{H}}$ using this work (see blue
  bands in Fig.~\ref{Fig:emcA3}) in combination with different
  theoretical predictions for $\mathcal{R}$ (see
  Eq.~\ref{Eq:f2nf2p_from_R}). The blue line here is the central value
  of the blue band in Fig.~\ref{Fig:f2nf2p}. See text for details.}
\label{Fig:F2nF2p_Systematics}
\end{figure*}

CT14 and CJ15 extracted parton distribution
functions rather than nucleon structure functions.  In order to
compare their results with our $F_2^n/F_2^p$ extraction, we
constructed the corresponding nucleon structure functions from their
individual parton distribution functions, accounting for valence
region corrections (higher-twist, target-mass) according to
Refs.~\cite{Accardi:2016qay,Accardi:2019pc}.  These corrections
largely cancel in the $F_2^n/F_2^p$ ratio.

The comparison with CJ15 is particularly interesting as that
extraction of $d(x_B)$ is predominantly constrained by the
D0 \ $W^\pm$ boson asymmetry data
\cite{D0:2014kma,Accardi:2016qay}, corresponding to $Q^2=m_W^2$. This
may indicate a tension between our low $Q^2$ results and results of
the CJ15 analysis of the D0 dataset at $x_B\ge0.6$.

We find that the $x_B \rightarrow 1$ limit of $F_2^n/F_2^p$ equals
$0.47 \pm 0.04$ for our Nuclear-DIS extraction.
Our results agree with predictions based on perturbative QCD~\cite{Farrar:1975yb} 
and the Dyson-Schwinger Equation (DSE)~\cite{Roberts:2013mja}
and disagree with  the Scalar Diquark model prediction~\cite{Close:1973xw, Carlitz:1975bg}.
This disagrees with the previous extractions (that apply nuclear corrections to the deuteron but do 
not consistently use data from heavier nuclei) that either could not discriminate among predictions, or
preferred the scalar diquark prediction. Our result is consistent with the upper edge of the CT14 extraction, 
which does not rely on nuclear corrections. However,
our $F_2^n/F_2^p$ has much smaller uncertainties which allow us to discriminate among models.

Thus, accounting for the modification of nucleons bound in deuterium increases $F_2^n$ at high-$x_B$.  
This was seen previously, see e.g. Ref.~\cite{Melnitchouk:1995fc, Yang:1998zb, Frankfurt88, hen11}.
However, the magnitude of this increase  at $x_B > 0.6$ is larger in our analysis as 
compared with those analyses that only use deuterium data. 
The high-$x_B$ disagreement between our nuclear DIS analysis and the analyses of Refs.~\cite{Dulat:2015mca,Accardi:2016qay,Arrington:2011qt} 
underscores the need for the forthcoming independent extraction by the MARATHON collaboration.
Below we present our predictions for their observables and quantify the model uncertainty associated with their $F_2^n/F_2^p$ extraction.

\section{$F_2^n/F_2^p$: Extraction from $A=3$ Mirror-Nuclei Data}

The MARATHON experiment recently measured DIS on $^2$H, $^3$H and
$^3$He.  They plan to independently extract $F_2^n/F_2^p$ from
$F_2^{^3\mathrm{He}}/F_2^{^3\mathrm{H}}$ using~\cite{MARATHON}:
\begin{eqnarray}
\begin{split}
\frac{F_2^n}{F_2^p} = \frac{2\mathcal{R} - F_2^{^3\mathrm{He}}/F_2^{^3\mathrm{H}}}{2 F_2^{^3\mathrm{He}}/F_2^{^3\mathrm{H}} - \mathcal{R}},
\label{Eq:f2nf2p_from_R}
\end{split}
\end{eqnarray}
where $\mathcal{R}$ is a theoretical correction factor which measures
the cancellation of nuclear effects in
$F_2^{^3\mathrm{He}}/F_2^{^3\mathrm{H}}$,
\begin{eqnarray}
\begin{split}
\mathcal{R} \equiv \frac{F_2^{^3\mathrm{He}}}{2F_2^p+F_2^n} \times \frac{F_2^p+2F_2^n}{F_2^{^3\mathrm{H}}}.
\label{Eq:R}
\end{split}
\end{eqnarray}
Since $^{3}$He and $^{3}$H should have similar nuclear effects
$\mathcal{R}$ should be close to 1.

We use our UMF to predict the expected DIS ratios for
$[F_2^{^3\mathrm{He}}/3]/[F_2^d/2]$,
$[F_2^{^3\mathrm{H}}/3]/[F_2^d/2]$, and
$F_2^{^3\mathrm{He}}/F_2^{^3\mathrm{H}}$ (see
Fig.~\ref{Fig:emcA3}). Since the $n_{SRC}^{^3\mathrm{H}}/n_{SRC}^d$
data are not yet published, we assumed $n_{SRC}^{^3\mathrm{H}}=
n_{SRC}^{^3\mathrm{He}}$.  Varying this by $\pm20\%$ changed our
results by less than $5\%$ at moderate and high-$x$, see online
supplementary materials.  

We compare our predictions for $[F_2^{^3\mathrm{He}}/3]/[F_2^d/2]$,
$[F_2^{^3\mathrm{H}}/3]/[F_2^d/2]$, and
$F_2^{^3\mathrm{He}}/F_2^{^3\mathrm{H}}$ with other models, shown as
colored lines in Fig.~\ref{Fig:emcA3}. Our prediction is overall
similar to that of Kulagin and Petti
(KP)~\cite{Kulagin:2019pc,Kulagin:2010gd}, though there are
differences at high $x_B$ in the $[F_2^{^3\mathrm{He}}/3]/[F_2^d/2]$,
and $F_2^{^3\mathrm{He}}/F_2^{^3\mathrm{H}}$ ratios. The Tropiano et
al. (TEMS) analysis~\cite{Tropiano:2019tems} combine the CJ15 global
PDF fits~\cite{Accardi:2016qay} and their off-shell correction in
deuterium, with additional fits to $[F_2^{^3\mathrm{He}}/3]/[F_2^d/2]$
data~\cite{Seely:2009gt}, to extract off-shell corrections in $A = 3$
nuclei. TEMS-CJ assumes fully isoscalar off-shell corrections.  In~\cite{Tropiano:2019tems}, fits
allowing non-isoscalar off-shell corrections were also performed,
which required an isoscalar correction as input.
TEMS-CJ$_\textrm{non-iso}$ uses the isoscalar correction from CJ15,
while TEMS-KP$_\textrm{non-iso}$ uses a different isoscalar
correction, developed by Kulagin and
Petti~\cite{Kulagin:2019pc,Kulagin:2010gd}. 
For $x_B > 0.6$, TEMS-CJ$_\textrm{non-iso}$ and TEMS-KP$_\textrm{non-iso}$ 
predictions~\cite{Tropiano:2019tems} individually disagree with our prediction of 
$F_2^{^3\mathrm{He}}/F_2^{^3\mathrm{H}}$. However, the spread of the two curves 
at $x_B > 0.6$ highlights the minimal sensitivity that $[F_2^{^3\mathrm{He}}/3]/[F_2^d/2]$ alone 
can provide to constraining non-isoscalar off-shell effects.
We agree with the isoscalar off-shell predictions of TEMS-CJ up to $x_B\sim0.5$. For 
$x_B<0.5$, even including uncertainty of TEMS-CJ$_\textrm{non-iso}$ and 
TEMS-KP$_\textrm{non-iso}$ (see supplementary materials), we predict a slightly higher ratio 
as compared to these two predictions.

We also studied the effect of different models of $\mathcal{R}$ on the
extractions of $F_2^n / F_2^p$ from
$F_2^{^3\mathrm{He}}/F_2^{^3\mathrm{H}}$.
Fig.~\ref{Fig:F2nF2p_Systematics} (left panel) shows several
theoretical predictions of $\mathcal{R}$. While individual models vary
by only a few percent, the choice of model can lead to significant, 
differences in the extracted $F_2^n / F_2^p$, especially at large
$x_B$.  Fig.~\ref{Fig:F2nF2p_Systematics} (right panel) shows $F_2^n /
F_2^p$ extracted using Eq.~\ref{Eq:f2nf2p_from_R}. Here we assume
$F_2^{^3\mathrm{He}}/F_2^{^3\mathrm{H}}$ from our Nuclear-DIS analysis
and then use various models of $\mathcal{R}$ to extract $F_2^n/F_2^p$, similar to 
the extraction MARATHON will perform with their measured $F_2^{^3\mathrm{He}}/F_2^{^3\mathrm{H}}$. 
While our prediction for $F_2^{^3\mathrm{He}}/F_2^{^3\mathrm{H}}$ is similar 
to that of KP (see Fig.~\ref{Fig:emcA3}), the differences at $x_B>0.7$ create 
large differences in $\mathcal{R}$, which cause a $\sim 10\%$ difference in the
extracted $F_2^n/F_2^p$.  The predictions of TEMS~\cite{Tropiano:2019tems} lead to larger
differences in $F_2^n/F_2^p$ and therefore even larger model uncertainties at
large-$x_B$~\cite{Sargsian:2001gu}. Performing the extraction of $F_2^n/F_2^p$ 
with different models for $F_2^{^3\mathrm{He}}/F_2^{^3\mathrm{H}}$ give similar uncertainty in 
$F_2^n/F_2^p$; see supplementary materials Fig 7.

Once the MARATHON $F_2^{^3\mathrm{He}}/F_2^{^3\mathrm{H}}$ data is
published, this model uncertainty could be reduced by iteratively
improving the extracted $F_2^n$ using Eqs.~\ref{Eq:f2nf2p_from_R}
and~\ref{Eq:R}~\cite{MARATHON}.  However, in this procedure, care
must be taken to ensure consistency with global nuclear DIS data,
as was done in our analysis.


\section{Conclusions}
Using Bayesian inference by means of a Hamiltonian Markov Chain Monte
Carlo, we extracted a nucleon universal modification function (UMF)
that is consistent with DIS measurements of nuclei from $A = 2$ to
208. We used it to correct Deuteron DIS data for bound-nucleon
structure-modification effects and to extract $F_2^n / F_2^p$ up to
$x_B \approx 0.9$.

The extracted $F_2^n / F_2^p$ ratio saturates at high-$x_B$ at a value
of $0.47 \pm 0.04$, which is consistent with perturbative QCD and DSE
predictions~\cite{Farrar:1975yb, Roberts:2013mja}, is lower than the
SU(6) symmetry prediction of 2/3~\cite{Close:1979bt}, and is
significantly greater than the Scalar Diquark model prediction of
1/4~\cite{Close:1973xw, Carlitz:1975bg}. Our Nuclear-DIS analysis
prediction also agrees with the most recent experimental extraction
by the BONuS experiment \cite{Baillie:2011za}.  The BONuS experiment
will take more data soon at higher energies and provide a more stringent
test of our predictions. The forthcoming parity-violating DIS program
using SoLID at Jefferson Lab will further probe $d/u$ directly using a
proton target~\cite{Solid:2014wp}.

We also used the UMF to predict the Tritium and $^3$He DIS cross section ratios, recently measured by the MARATHON experiment \cite{MARATHON}, and to estimate the nuclear correction function $\mathcal{R}$ that they plan to use to extract $F_2^n / F_2^p$ from their data. We showed that different models of $\mathcal{R}$ lead to non-negligible model uncertainty in the planned extraction of $F_2^n / F_2^p$.


\begin{acknowledgments}
We thank C. Keppel, W. Melnitchouk, and N. Sato for useful discussions.
This work was supported by the U.S. Department of Energy, Office of Science, Office of Nuclear Physics under Award Numbers DE-FG02-94ER40818, DE-FG02-96ER-40960, DE-FG02-93ER40771, and DE-AC05-06OR23177 under which Jefferson Science Associates operates the Thomas Jefferson National Accelerator Facility, the Pazy foundation, and the Israeli Science Foundation (Israel) under Grants Nos. 136/12 and 1334/16. 
\end{acknowledgments}

\bibliography{EMC_Barak_bib}

\begin{thebibliography}{49}
\expandafter\ifx\csname natexlab\endcsname\relax\def\natexlab#1{#1}\fi
\expandafter\ifx\csname bibnamefont\endcsname\relax
  \def\bibnamefont#1{#1}\fi
\expandafter\ifx\csname bibfnamefont\endcsname\relax
  \def\bibfnamefont#1{#1}\fi
\expandafter\ifx\csname citenamefont\endcsname\relax
  \def\citenamefont#1{#1}\fi
\expandafter\ifx\csname url\endcsname\relax
  \def\url#1{\texttt{#1}}\fi
\expandafter\ifx\csname urlprefix\endcsname\relax\def\urlprefix{URL }\fi
\providecommand{\bibinfo}[2]{#2}
\providecommand{\eprint}[2][]{\url{#2}}

\bibitem[{\citenamefont{Bashir et~al.}(2012)\citenamefont{Bashir, Chang, Cloet,
  El-Bennich, Liu, Roberts, and Tandy}}]{Bashir:2012}
\bibinfo{author}{\bibfnamefont{A.}~\bibnamefont{Bashir}},
  \bibinfo{author}{\bibfnamefont{L.}~\bibnamefont{Chang}},
  \bibinfo{author}{\bibfnamefont{I.}~\bibnamefont{Cloet}},
  \bibinfo{author}{\bibfnamefont{B.}~\bibnamefont{El-Bennich}},
  \bibinfo{author}{\bibfnamefont{Y.}~\bibnamefont{Liu}},
  \bibinfo{author}{\bibfnamefont{C.}~\bibnamefont{Roberts}}, \bibnamefont{and}
  \bibinfo{author}{\bibfnamefont{P.}~\bibnamefont{Tandy}},
  \bibinfo{journal}{Commun. Theor. Phys.} \textbf{\bibinfo{volume}{58}},
  \bibinfo{pages}{79} (\bibinfo{year}{2012}), \eprint{1201.3366}.

\bibitem[{\citenamefont{Roberts et~al.}(2013)\citenamefont{Roberts, Holt, and
  Schmidt}}]{Roberts:2013mja}
\bibinfo{author}{\bibfnamefont{C.~D.} \bibnamefont{Roberts}},
  \bibinfo{author}{\bibfnamefont{R.~J.} \bibnamefont{Holt}}, \bibnamefont{and}
  \bibinfo{author}{\bibfnamefont{S.~M.} \bibnamefont{Schmidt}},
  \bibinfo{journal}{Phys. Lett.} \textbf{\bibinfo{volume}{B727}},
  \bibinfo{pages}{249} (\bibinfo{year}{2013}), \eprint{1308.1236}.

\bibitem[{\citenamefont{Farrar and Jackson}(1975)}]{Farrar:1975yb}
\bibinfo{author}{\bibfnamefont{G.~R.} \bibnamefont{Farrar}} \bibnamefont{and}
  \bibinfo{author}{\bibfnamefont{D.~R.} \bibnamefont{Jackson}},
  \bibinfo{journal}{Phys. Rev. Lett.} \textbf{\bibinfo{volume}{35}},
  \bibinfo{pages}{1416} (\bibinfo{year}{1975}).

\bibitem[{\citenamefont{Dulat et~al.}(2016)\citenamefont{Dulat, Hou, Gao,
  Guzzi, Huston, Nadolsky, Pumplin, Schmidt, Stump, and Yuan}}]{Dulat:2015mca}
\bibinfo{author}{\bibfnamefont{S.}~\bibnamefont{Dulat}},
  \bibinfo{author}{\bibfnamefont{T.-J.} \bibnamefont{Hou}},
  \bibinfo{author}{\bibfnamefont{J.}~\bibnamefont{Gao}},
  \bibinfo{author}{\bibfnamefont{M.}~\bibnamefont{Guzzi}},
  \bibinfo{author}{\bibfnamefont{J.}~\bibnamefont{Huston}},
  \bibinfo{author}{\bibfnamefont{P.}~\bibnamefont{Nadolsky}},
  \bibinfo{author}{\bibfnamefont{J.}~\bibnamefont{Pumplin}},
  \bibinfo{author}{\bibfnamefont{C.}~\bibnamefont{Schmidt}},
  \bibinfo{author}{\bibfnamefont{D.}~\bibnamefont{Stump}}, \bibnamefont{and}
  \bibinfo{author}{\bibfnamefont{C.~P.} \bibnamefont{Yuan}},
  \bibinfo{journal}{Phys. Rev.} \textbf{\bibinfo{volume}{D93}},
  \bibinfo{pages}{033006} (\bibinfo{year}{2016}), \eprint{1506.07443}.

\bibitem[{\citenamefont{Accardi et~al.}(2016)\citenamefont{Accardi, Brady,
  Melnitchouk, Owens, and Sato}}]{Accardi:2016qay}
\bibinfo{author}{\bibfnamefont{A.}~\bibnamefont{Accardi}},
  \bibinfo{author}{\bibfnamefont{L.~T.} \bibnamefont{Brady}},
  \bibinfo{author}{\bibfnamefont{W.}~\bibnamefont{Melnitchouk}},
  \bibinfo{author}{\bibfnamefont{J.~F.} \bibnamefont{Owens}}, \bibnamefont{and}
  \bibinfo{author}{\bibfnamefont{N.}~\bibnamefont{Sato}},
  \bibinfo{journal}{Phys. Rev.} \textbf{\bibinfo{volume}{D93}},
  \bibinfo{pages}{114017} (\bibinfo{year}{2016}), \eprint{1602.03154}.

\bibitem[{\citenamefont{Arrington et~al.}(2012)\citenamefont{Arrington, Rubin,
  and Melnitchouk}}]{Arrington:2011qt}
\bibinfo{author}{\bibfnamefont{J.}~\bibnamefont{Arrington}},
  \bibinfo{author}{\bibfnamefont{J.~G.} \bibnamefont{Rubin}}, \bibnamefont{and}
  \bibinfo{author}{\bibfnamefont{W.}~\bibnamefont{Melnitchouk}},
  \bibinfo{journal}{Phys. Rev. Lett.} \textbf{\bibinfo{volume}{108}},
  \bibinfo{pages}{252001} (\bibinfo{year}{2012}), \eprint{1110.3362}.

\bibitem[{\citenamefont{Close}(1973)}]{Close:1973xw}
\bibinfo{author}{\bibfnamefont{F.~E.} \bibnamefont{Close}},
  \bibinfo{journal}{Phys. Lett.} \textbf{\bibinfo{volume}{43B}},
  \bibinfo{pages}{422} (\bibinfo{year}{1973}).

\bibitem[{\citenamefont{Carlitz}(1975)}]{Carlitz:1975bg}
\bibinfo{author}{\bibfnamefont{R.~D.} \bibnamefont{Carlitz}},
  \bibinfo{journal}{Phys. Lett.} \textbf{\bibinfo{volume}{58B}},
  \bibinfo{pages}{345} (\bibinfo{year}{1975}).

\bibitem[{\citenamefont{Petratos et~al.}(2010)}]{MARATHON}
\bibinfo{author}{\bibfnamefont{G.~G.} \bibnamefont{Petratos}}
  \bibnamefont{et~al.}, \bibinfo{journal}{Jefferson Lab PAC37 Proposal}
  (\bibinfo{year}{2010}), \bibinfo{note}{experiment E12-10-103}.

\bibitem[{\citenamefont{Hen et~al.}(2017)\citenamefont{Hen, Miller, Piasetzky,
  and Weinstein}}]{Hen:2016kwk}
\bibinfo{author}{\bibfnamefont{O.}~\bibnamefont{Hen}},
  \bibinfo{author}{\bibfnamefont{G.~A.} \bibnamefont{Miller}},
  \bibinfo{author}{\bibfnamefont{E.}~\bibnamefont{Piasetzky}},
  \bibnamefont{and} \bibinfo{author}{\bibfnamefont{L.~B.}
  \bibnamefont{Weinstein}}, \bibinfo{journal}{Rev. Mod. Phys.}
  \textbf{\bibinfo{volume}{89}}, \bibinfo{pages}{045002}
  (\bibinfo{year}{2017}).

\bibitem[{\citenamefont{Weinstein et~al.}(2011)\citenamefont{Weinstein,
  Piasetzky, Higinbotham, Gomez, Hen, and Shneor}}]{weinstein11}
\bibinfo{author}{\bibfnamefont{L.~B.} \bibnamefont{Weinstein}},
  \bibinfo{author}{\bibfnamefont{E.}~\bibnamefont{Piasetzky}},
  \bibinfo{author}{\bibfnamefont{D.~W.} \bibnamefont{Higinbotham}},
  \bibinfo{author}{\bibfnamefont{J.}~\bibnamefont{Gomez}},
  \bibinfo{author}{\bibfnamefont{O.}~\bibnamefont{Hen}}, \bibnamefont{and}
  \bibinfo{author}{\bibfnamefont{R.}~\bibnamefont{Shneor}},
  \bibinfo{journal}{Phys. Rev. Lett.} \textbf{\bibinfo{volume}{106}},
  \bibinfo{pages}{052301} (\bibinfo{year}{2011}).

\bibitem[{\citenamefont{Hen et~al.}(2012)\citenamefont{Hen, Piasetzky, and
  Weinstein}}]{Hen12}
\bibinfo{author}{\bibfnamefont{O.}~\bibnamefont{Hen}},
  \bibinfo{author}{\bibfnamefont{E.}~\bibnamefont{Piasetzky}},
  \bibnamefont{and} \bibinfo{author}{\bibfnamefont{L.~B.}
  \bibnamefont{Weinstein}}, \bibinfo{journal}{Phys. Rev. C}
  \textbf{\bibinfo{volume}{85}}, \bibinfo{pages}{047301}
  (\bibinfo{year}{2012}).

\bibitem[{\citenamefont{Hen et~al.}(2013)\citenamefont{Hen, Higinbotham,
  Miller, Piasetzky, and Weinstein}}]{Hen:2013oha}
\bibinfo{author}{\bibfnamefont{O.}~\bibnamefont{Hen}},
  \bibinfo{author}{\bibfnamefont{D.~W.} \bibnamefont{Higinbotham}},
  \bibinfo{author}{\bibfnamefont{G.~A.} \bibnamefont{Miller}},
  \bibinfo{author}{\bibfnamefont{E.}~\bibnamefont{Piasetzky}},
  \bibnamefont{and} \bibinfo{author}{\bibfnamefont{L.~B.}
  \bibnamefont{Weinstein}}, \bibinfo{journal}{Int. J. Mod. Phys.}
  \textbf{\bibinfo{volume}{E22}}, \bibinfo{pages}{1330017}
  (\bibinfo{year}{2013}), \eprint{1304.2813}.

\bibitem[{\citenamefont{Schmookler et~al.}(2019)}]{Schmookler:2019nvf}
\bibinfo{author}{\bibfnamefont{B.}~\bibnamefont{Schmookler}}
  \bibnamefont{et~al.} (\bibinfo{collaboration}{CLAS}),
  \bibinfo{journal}{Nature} \textbf{\bibinfo{volume}{566}},
  \bibinfo{pages}{354} (\bibinfo{year}{2019}).

\bibitem[{\citenamefont{Frankfurt and Strikman}(1988)}]{Frankfurt88}
\bibinfo{author}{\bibfnamefont{L.}~\bibnamefont{Frankfurt}} \bibnamefont{and}
  \bibinfo{author}{\bibfnamefont{M.}~\bibnamefont{Strikman}},
  \bibinfo{journal}{Phys. Rep.} \textbf{\bibinfo{volume}{160}},
  \bibinfo{pages}{235 } (\bibinfo{year}{1988}).

\bibitem[{\citenamefont{Piasetzky et~al.}(2006)\citenamefont{Piasetzky,
  Sargsian, Frankfurt, Strikman, and Watson}}]{piasetzky06}
\bibinfo{author}{\bibfnamefont{E.}~\bibnamefont{Piasetzky}},
  \bibinfo{author}{\bibfnamefont{M.}~\bibnamefont{Sargsian}},
  \bibinfo{author}{\bibfnamefont{L.}~\bibnamefont{Frankfurt}},
  \bibinfo{author}{\bibfnamefont{M.}~\bibnamefont{Strikman}}, \bibnamefont{and}
  \bibinfo{author}{\bibfnamefont{J.~W.} \bibnamefont{Watson}},
  \bibinfo{journal}{Phys. Rev. Lett.} \textbf{\bibinfo{volume}{97}},
  \bibinfo{pages}{162504} (\bibinfo{year}{2006}).

\bibitem[{\citenamefont{Subedi et~al.}(2008)}]{subedi08}
\bibinfo{author}{\bibfnamefont{R.}~\bibnamefont{Subedi}} \bibnamefont{et~al.},
  \bibinfo{journal}{Science} \textbf{\bibinfo{volume}{320}},
  \bibinfo{pages}{1476} (\bibinfo{year}{2008}).

\bibitem[{\citenamefont{Korover et~al.}(2014)\citenamefont{Korover, Muangma,
  Hen et~al.}}]{korover14}
\bibinfo{author}{\bibfnamefont{I.}~\bibnamefont{Korover}},
  \bibinfo{author}{\bibfnamefont{N.}~\bibnamefont{Muangma}},
  \bibinfo{author}{\bibfnamefont{O.}~\bibnamefont{Hen}}, \bibnamefont{et~al.},
  \bibinfo{journal}{Phys.Rev.Lett.} \textbf{\bibinfo{volume}{113}},
  \bibinfo{pages}{022501} (\bibinfo{year}{2014}).

\bibitem[{\citenamefont{Hen et~al.}(2014)}]{hen14}
\bibinfo{author}{\bibfnamefont{O.}~\bibnamefont{Hen}} \bibnamefont{et~al.}
  (\bibinfo{collaboration}{CLAS Collaboration}), \bibinfo{journal}{Science}
  \textbf{\bibinfo{volume}{346}}, \bibinfo{pages}{614} (\bibinfo{year}{2014}).

\bibitem[{\citenamefont{Duer et~al.}(2018)}]{duer18}
\bibinfo{author}{\bibfnamefont{M.}~\bibnamefont{Duer}} \bibnamefont{et~al.}
  (\bibinfo{collaboration}{CLAS}), \bibinfo{journal}{Nature}
  \textbf{\bibinfo{volume}{560}}, \bibinfo{pages}{617} (\bibinfo{year}{2018}).

\bibitem[{\citenamefont{Duer et~al.}(2019)}]{Duer:2018sxh}
\bibinfo{author}{\bibfnamefont{M.}~\bibnamefont{Duer}} \bibnamefont{et~al.}
  (\bibinfo{collaboration}{CLAS}), \bibinfo{journal}{Phys. Rev. Lett.}
  \textbf{\bibinfo{volume}{122}}, \bibinfo{pages}{172502}
  (\bibinfo{year}{2019}), \eprint{1810.05343}.

\bibitem[{\citenamefont{Tang et~al.}(2003)}]{tang03}
\bibinfo{author}{\bibfnamefont{A.}~\bibnamefont{Tang}} \bibnamefont{et~al.},
  \bibinfo{journal}{Phys. Rev. Lett.} \textbf{\bibinfo{volume}{90}},
  \bibinfo{pages}{042301} (\bibinfo{year}{2003}).

\bibitem[{\citenamefont{Shneor et~al.}(2007)}]{shneor07}
\bibinfo{author}{\bibfnamefont{R.}~\bibnamefont{Shneor}} \bibnamefont{et~al.},
  \bibinfo{journal}{Phys. Rev. Lett.} \textbf{\bibinfo{volume}{99}},
  \bibinfo{eid}{072501} (\bibinfo{year}{2007}).

\bibitem[{\citenamefont{Cohen et~al.}(2018)}]{Cohen:2018gzh}
\bibinfo{author}{\bibfnamefont{E.~O.} \bibnamefont{Cohen}} \bibnamefont{et~al.}
  (\bibinfo{collaboration}{CLAS}), \bibinfo{journal}{Phys. Rev. Lett.}
  \textbf{\bibinfo{volume}{121}}, \bibinfo{pages}{092501}
  (\bibinfo{year}{2018}), \eprint{1805.01981}.

\bibitem[{\citenamefont{Sargsian et~al.}(2003)}]{Sargsian02}
\bibinfo{author}{\bibfnamefont{M.~M.} \bibnamefont{Sargsian}}
  \bibnamefont{et~al.}, \bibinfo{journal}{J. Phys.}
  \textbf{\bibinfo{volume}{G29}}, \bibinfo{pages}{R1} (\bibinfo{year}{2003}).

\bibitem[{\citenamefont{Melnitchouk et~al.}(1997)\citenamefont{Melnitchouk,
  Sargsian, and Strikman}}]{Melnitchouk:1997pemc}
\bibinfo{author}{\bibfnamefont{W.}~\bibnamefont{Melnitchouk}},
  \bibinfo{author}{\bibfnamefont{M.}~\bibnamefont{Sargsian}}, \bibnamefont{and}
  \bibinfo{author}{\bibfnamefont{M.}~\bibnamefont{Strikman}},
  \bibinfo{journal}{Z. Phys.} \textbf{\bibinfo{volume}{A359}},
  \bibinfo{pages}{99} (\bibinfo{year}{1997}).

\bibitem[{\citenamefont{Virchaux and Milsztajn}(1992)}]{Virchaux:1991jc}
\bibinfo{author}{\bibfnamefont{M.}~\bibnamefont{Virchaux}} \bibnamefont{and}
  \bibinfo{author}{\bibfnamefont{A.}~\bibnamefont{Milsztajn}},
  \bibinfo{journal}{Phys. Lett.} \textbf{\bibinfo{volume}{B274}},
  \bibinfo{pages}{221} (\bibinfo{year}{1992}).

\bibitem[{\citenamefont{Carpenter et~al.}(2017)\citenamefont{Carpenter, Gelman,
  Hoffman, Lee, Goodrich, Betancourt, Brubaker, Guo, Li, and
  Riddell}}]{JSSv076i01}
\bibinfo{author}{\bibfnamefont{B.}~\bibnamefont{Carpenter}},
  \bibinfo{author}{\bibfnamefont{A.}~\bibnamefont{Gelman}},
  \bibinfo{author}{\bibfnamefont{M.}~\bibnamefont{Hoffman}},
  \bibinfo{author}{\bibfnamefont{D.}~\bibnamefont{Lee}},
  \bibinfo{author}{\bibfnamefont{B.}~\bibnamefont{Goodrich}},
  \bibinfo{author}{\bibfnamefont{M.}~\bibnamefont{Betancourt}},
  \bibinfo{author}{\bibfnamefont{M.}~\bibnamefont{Brubaker}},
  \bibinfo{author}{\bibfnamefont{J.}~\bibnamefont{Guo}},
  \bibinfo{author}{\bibfnamefont{P.}~\bibnamefont{Li}}, \bibnamefont{and}
  \bibinfo{author}{\bibfnamefont{A.}~\bibnamefont{Riddell}},
  \bibinfo{journal}{Journal of Statistical Software, Articles}
  \textbf{\bibinfo{volume}{76}}, \bibinfo{pages}{1} (\bibinfo{year}{2017}),
  ISSN \bibinfo{issn}{1548-7660},
  \urlprefix\url{https://www.jstatsoft.org/v076/i01}.

\bibitem[{\citenamefont{Team}(2018)}]{pystan}
\bibinfo{author}{\bibfnamefont{S.~D.} \bibnamefont{Team}},
  \emph{\bibinfo{title}{Pystan: the python interface to stan, version
  2.17.1.0}}, \bibinfo{howpublished}{\url{http://mc-stan.org}}
  (\bibinfo{year}{2018}).

\bibitem[{\citenamefont{Gomez et~al.}(1994)}]{Gomez94}
\bibinfo{author}{\bibfnamefont{J.}~\bibnamefont{Gomez}} \bibnamefont{et~al.},
  \bibinfo{journal}{Phys. Rev. D} \textbf{\bibinfo{volume}{49}},
  \bibinfo{pages}{4348} (\bibinfo{year}{1994}).

\bibitem[{\citenamefont{Seely et~al.}(2009)}]{Seely:2009gt}
\bibinfo{author}{\bibfnamefont{J.}~\bibnamefont{Seely}} \bibnamefont{et~al.},
  \bibinfo{journal}{Phys. Rev. Lett.} \textbf{\bibinfo{volume}{103}},
  \bibinfo{pages}{202301} (\bibinfo{year}{2009}).

\bibitem[{\citenamefont{Frankfurt et~al.}(1993)\citenamefont{Frankfurt,
  Strikman, Day, and Sargsyan}}]{frankfurt93}
\bibinfo{author}{\bibfnamefont{L.}~\bibnamefont{Frankfurt}},
  \bibinfo{author}{\bibfnamefont{M.}~\bibnamefont{Strikman}},
  \bibinfo{author}{\bibfnamefont{D.}~\bibnamefont{Day}}, \bibnamefont{and}
  \bibinfo{author}{\bibfnamefont{M.}~\bibnamefont{Sargsyan}},
  \bibinfo{journal}{Phys. Rev. C} \textbf{\bibinfo{volume}{48}},
  \bibinfo{pages}{2451} (\bibinfo{year}{1993}).

\bibitem[{\citenamefont{Egiyan et~al.}(2003)}]{egiyan02}
\bibinfo{author}{\bibfnamefont{K.}~\bibnamefont{Egiyan}} \bibnamefont{et~al.}
  (\bibinfo{collaboration}{CLAS Collaboration}), \bibinfo{journal}{Phys. Rev.
  C} \textbf{\bibinfo{volume}{68}}, \bibinfo{pages}{014313}
  (\bibinfo{year}{2003}).

\bibitem[{\citenamefont{Egiyan et~al.}(2006)}]{egiyan06}
\bibinfo{author}{\bibfnamefont{K.}~\bibnamefont{Egiyan}} \bibnamefont{et~al.}
  (\bibinfo{collaboration}{CLAS Collaboration}), \bibinfo{journal}{Phys. Rev.
  Lett.} \textbf{\bibinfo{volume}{96}}, \bibinfo{pages}{082501}
  (\bibinfo{year}{2006}).

\bibitem[{\citenamefont{Fomin et~al.}(2012)}]{fomin12}
\bibinfo{author}{\bibfnamefont{N.}~\bibnamefont{Fomin}} \bibnamefont{et~al.},
  \bibinfo{journal}{Phys. Rev. Lett.} \textbf{\bibinfo{volume}{108}},
  \bibinfo{pages}{092502} (\bibinfo{year}{2012}).

\bibitem[{\citenamefont{Arrington et~al.}(2009)\citenamefont{Arrington,
  Coester, Holt, and Lee}}]{Arrington:2008zh}
\bibinfo{author}{\bibfnamefont{J.}~\bibnamefont{Arrington}},
  \bibinfo{author}{\bibfnamefont{F.}~\bibnamefont{Coester}},
  \bibinfo{author}{\bibfnamefont{R.~J.} \bibnamefont{Holt}}, \bibnamefont{and}
  \bibinfo{author}{\bibfnamefont{T.~S.~H.} \bibnamefont{Lee}},
  \bibinfo{journal}{J. Phys.} \textbf{\bibinfo{volume}{G36}},
  \bibinfo{pages}{025005} (\bibinfo{year}{2009}), \eprint{0805.3116}.

\bibitem[{\citenamefont{Baillie et~al.}(2012)}]{Baillie:2011za}
\bibinfo{author}{\bibfnamefont{N.}~\bibnamefont{Baillie}} \bibnamefont{et~al.}
  (\bibinfo{collaboration}{CLAS}), \bibinfo{journal}{Phys. Rev. Lett.}
  \textbf{\bibinfo{volume}{108}}, \bibinfo{pages}{142001}
  (\bibinfo{year}{2012}), \bibinfo{note}{[Erratum: Phys. Rev.
  Lett.108,199902(2012)]}, \eprint{1110.2770}.

\bibitem[{\citenamefont{Abazov et~al.}(2015)}]{D0:2014kma}
\bibinfo{author}{\bibfnamefont{V.~M.} \bibnamefont{Abazov}}
  \bibnamefont{et~al.} (\bibinfo{collaboration}{D0}), \bibinfo{journal}{Phys.
  Rev.} \textbf{\bibinfo{volume}{D91}}, \bibinfo{pages}{032007}
  (\bibinfo{year}{2015}), \bibinfo{note}{[Erratum: Phys.
  Rev.D91,no.7,079901(2015)]}, \eprint{1412.2862}.

\bibitem[{\citenamefont{Close}(1979)}]{Close:1979bt}
\bibinfo{author}{\bibfnamefont{F.~E.} \bibnamefont{Close}},
  \emph{\bibinfo{title}{{An Introduction to Quarks and Partons}}}
  (\bibinfo{publisher}{Academic Press}, \bibinfo{address}{London},
  \bibinfo{year}{1979}).

\bibitem[{\citenamefont{Tropiano et~al.}(2019)\citenamefont{Tropiano, Ethier,
  Melnitchouk, and Sato}}]{Tropiano:2019tems}
\bibinfo{author}{\bibfnamefont{A.}~\bibnamefont{Tropiano}},
  \bibinfo{author}{\bibfnamefont{J.}~\bibnamefont{Ethier}},
  \bibinfo{author}{\bibfnamefont{W.}~\bibnamefont{Melnitchouk}},
  \bibnamefont{and} \bibinfo{author}{\bibfnamefont{N.}~\bibnamefont{Sato}},
  \bibinfo{journal}{Phys. Rev.} \textbf{\bibinfo{volume}{C99}},
  \bibinfo{pages}{035201} (\bibinfo{year}{2019}), \eprint{1811.07668}.

\bibitem[{\citenamefont{Kulagin and Petti}(private
  communication)}]{Kulagin:2019pc}
\bibinfo{author}{\bibfnamefont{S.~A.} \bibnamefont{Kulagin}} \bibnamefont{and}
  \bibinfo{author}{\bibfnamefont{R.}~\bibnamefont{Petti}}
  (\bibinfo{year}{private communication}).

\bibitem[{\citenamefont{Kulagin and Petti}(2010)}]{Kulagin:2010gd}
\bibinfo{author}{\bibfnamefont{S.~A.} \bibnamefont{Kulagin}} \bibnamefont{and}
  \bibinfo{author}{\bibfnamefont{R.}~\bibnamefont{Petti}},
  \bibinfo{journal}{Phys. Rev.} \textbf{\bibinfo{volume}{C82}},
  \bibinfo{pages}{054614} (\bibinfo{year}{2010}), \eprint{1004.3062}.

\bibitem[{\citenamefont{Afnan et~al.}(2003)\citenamefont{Afnan, Bissey, Gomez,
  Katramatou, Liuti, Melnitchouk, Petratos, and Thomas}}]{Afnan:2003vh}
\bibinfo{author}{\bibfnamefont{I.~R.} \bibnamefont{Afnan}},
  \bibinfo{author}{\bibfnamefont{F.~R.~P.} \bibnamefont{Bissey}},
  \bibinfo{author}{\bibfnamefont{J.}~\bibnamefont{Gomez}},
  \bibinfo{author}{\bibfnamefont{A.~T.} \bibnamefont{Katramatou}},
  \bibinfo{author}{\bibfnamefont{S.}~\bibnamefont{Liuti}},
  \bibinfo{author}{\bibfnamefont{W.}~\bibnamefont{Melnitchouk}},
  \bibinfo{author}{\bibfnamefont{G.~G.} \bibnamefont{Petratos}},
  \bibnamefont{and} \bibinfo{author}{\bibfnamefont{A.~W.}
  \bibnamefont{Thomas}}, \bibinfo{journal}{Phys. Rev.}
  \textbf{\bibinfo{volume}{C68}}, \bibinfo{pages}{035201}
  (\bibinfo{year}{2003}), \eprint{nucl-th/0306054}.

\bibitem[{\citenamefont{Accardi}(private communication)}]{Accardi:2019pc}
\bibinfo{author}{\bibfnamefont{A.}~\bibnamefont{Accardi}}
  (\bibinfo{year}{private communication}).

\bibitem[{\citenamefont{Melnitchouk and Thomas}(1996)}]{Melnitchouk:1995fc}
\bibinfo{author}{\bibfnamefont{W.}~\bibnamefont{Melnitchouk}} \bibnamefont{and}
  \bibinfo{author}{\bibfnamefont{A.~W.} \bibnamefont{Thomas}},
  \bibinfo{journal}{Phys. Lett.} \textbf{\bibinfo{volume}{B377}},
  \bibinfo{pages}{11} (\bibinfo{year}{1996}), \eprint{nucl-th/9602038}.

\bibitem[{\citenamefont{Yang and Bodek}(1999)}]{Yang:1998zb}
\bibinfo{author}{\bibfnamefont{U.-K.} \bibnamefont{Yang}} \bibnamefont{and}
  \bibinfo{author}{\bibfnamefont{A.}~\bibnamefont{Bodek}},
  \bibinfo{journal}{Phys. Rev. Lett.} \textbf{\bibinfo{volume}{82}},
  \bibinfo{pages}{2467} (\bibinfo{year}{1999}), \eprint{hep-ph/9809480}.

\bibitem[{\citenamefont{Hen et~al.}(2011)\citenamefont{Hen, Accardi,
  Melnitchouk, and Piasetzky}}]{hen11}
\bibinfo{author}{\bibfnamefont{O.}~\bibnamefont{Hen}},
  \bibinfo{author}{\bibfnamefont{A.}~\bibnamefont{Accardi}},
  \bibinfo{author}{\bibfnamefont{W.}~\bibnamefont{Melnitchouk}},
  \bibnamefont{and}
  \bibinfo{author}{\bibfnamefont{E.}~\bibnamefont{Piasetzky}},
  \bibinfo{journal}{Phys. Rev. D} \textbf{\bibinfo{volume}{84}},
  \bibinfo{pages}{117501} (\bibinfo{year}{2011}).

\bibitem[{\citenamefont{Sargsian et~al.}(2002)\citenamefont{Sargsian, Simula,
  and Strikman}}]{Sargsian:2001gu}
\bibinfo{author}{\bibfnamefont{M.~M.} \bibnamefont{Sargsian}},
  \bibinfo{author}{\bibfnamefont{S.}~\bibnamefont{Simula}}, \bibnamefont{and}
  \bibinfo{author}{\bibfnamefont{M.~I.} \bibnamefont{Strikman}},
  \bibinfo{journal}{Phys. Rev.} \textbf{\bibinfo{volume}{C66}},
  \bibinfo{pages}{024001} (\bibinfo{year}{2002}), \eprint{nucl-th/0105052}.

\bibitem[{\citenamefont{{SoLID Collaboration}}(2014)}]{Solid:2014wp}
\bibinfo{author}{\bibnamefont{{SoLID Collaboration}}},
  \bibinfo{howpublished}{\url{https://hallaweb.jlab.org/12GeV/SoLID/download/doc/SoLIDWhitePaper-Sep9-2014.pdfu}}
  (\bibinfo{year}{2014}).

\end{thebibliography}

\end{document}